\documentstyle[sprocl]{article}

\def\be{\begin{equation}}
\def\ee{\end{equation}}
\def\bea{\begin{eqnarray}}
\def\eea{\end{eqnarray}}
\begin{document}

\title{STRONG AND WEAK LENSING CONSTRAINTS \\ON GALAXY MASS DISTRIBUTION}

\author{Jean-Paul KNEIB}

\address{Observatoire Midi-Pyr\'en\'ees, 14 Av. E.Belin,
31400 Toulouse, France\\E-mail: jean-paul.kneib@ast.obs-mip.fr} 

\maketitle\abstracts{Gravitational Lensing is a {\sl unique} tool
to constrain the mass distribution of collapsed structures, this is
particularly true for galaxies, either on a case by case basis using
multiple images of background sources (such as quasars), 
or statistically using the so called {\it galaxy-galaxy lensing} technique. 
First, I will present the lensing theory, and then discuss the various
methods applied to current observations. Finally, I will review
the bright future prospects of galaxy lensing that will benefit
of the development of high resolution, large, wide and deep (lensing) surveys.
}

\section{The Theory: \hfill -- What Do We Expect --}

Although, Astrophysics is generally a question of detecting photons 
and making sense of them using various physical theories, 
Gravitational Lensing (GL) is generally a question of understanding the 
light paths of these photons which then allow to probe the 
intervening mass distribution. Thus GL must be seen as a useful tool
(in a similar way as stellar dynamics) to probe the mass distribution
of galaxies - objects of interest in this Conference.

\subsection{Useful Gravitational Lensing Equations}\label{subsec:GL}

A lens operates a transformation from Source plane to Image plane
that is merely a simple 2D-mapping that can be describe by 3 equations 
relating the Source and Image properties:
\begin{enumerate}
\item Position:
\vspace*{-4mm}
\begin{equation}
\vec{\theta_S}= \vec{\theta_I} - \vec{\nabla}\varphi(\vec{\theta_I})
\end{equation}
this equation give the position of the source $\vec\theta_S$ for an 
image at position $\vec{\theta_I}$;
$\varphi$ is the lensing potential that relates to the
projected Newtonian potential by: 
$\varphi = {2\over c^2}{D_{OS}\over D_{LS}} \phi^{2D}$.
Furthermore, as the mapping is purely geometrical, the surface brightness of the
objects are conserved through the equation: 
$S(\vec{\theta_I})=S(\vec{\theta_S})$.

\item Shape:
\vspace*{-2mm}
\begin{equation}
{d\vec{\theta_S}\over d\vec{\theta_I}} = A^{-1} = 
\left(
\begin{array}{cc}
1-\partial_{xx}\varphi(\vec{\theta_I}) & -\partial_{xy}\varphi(\vec{\theta_I})\\
-\partial_{yx}\varphi(\vec{\theta_I}) & 1-\partial_{yy}\varphi(\vec{\theta_I})\\
\end{array}
\right)
\equiv
\left(
\begin{array}{cc}
1-\kappa -\gamma_1 & -\gamma_2\\
-\gamma_2 & 1-\kappa +\gamma_1\\
\end{array}
\right)
\end{equation}
This equation defines the inverse of the amplification matrix $A^{-1}$ 
at the image position $\vec{\theta_I}$.
The image magnification $\mu$ is defined as:
$ \mu^{-1}=\det(A^{-1})$ and  the 
reduced shear (distortion induced by the mass distribution) is
$\vec{g} = {\vec{\gamma}\over 1-\kappa}$ where $2\kappa = \Sigma/\Sigma_{crit}$,
with $\Sigma_{crit}= {c^2\over 4G}{D_{OS} D_{OL}
  \over D_{LS} }$ the {\it critical density}.
We also define the absolute shear $\bar\gamma =
{D_{OS}\over D_{LS}} \gamma$
which is independent on the redshift  of the lens and the source and only
depends on the mass distribution.

\item Time:
\vspace*{-4mm}
\begin{equation}
t_a \propto {1 \over 2}(\vec\theta_S - \vec\theta_I)^2 - \varphi(\vec{\theta_I})
\end{equation}
This equation defines the arrival time of an image at position $\vec{\theta_I}$.
The difference between two images gives the time-delay. Originally
it was foreseen to use the time-delay to constrain the Hubble constant 
$H_0$, but it does strongly depend on the exact mass profile and distribution.
Therefore, assuming a reasonable value for $H_0$, the measure of the time-delay
can put strong constraints on the mass distribution, in particular
in constraining the total convergence  of the lens.
\end{enumerate}

We will concentrate here on the determination of the mass distribution, thus
assuming that we have measured the source and lens redshifts as well
as determined/fixed the cosmology.

\subsection{Case of a Circular Mass Distribution}\label{subsec:CMD}

All 3 equations depend on the lensing potential $\varphi$
either by its gradient (position), its second derivatives (shape/amplification)
or its value (time-delay). For a circular mass distribution, 
it is easy to show that
$\partial_r\varphi ={m(r)\over r}$, and that all lensing 
equations can be written as functions of ${m(r)\over r}\propto \bar\Sigma(<r)$.
Note that constraints are absolute in the case of multiple images
but are only relative in terms of shear (galaxy-galaxy lensing) as in this
case ${m(r)\over r}$ can be expressed as an integral for which the
limits are not always well defined [this is also true in cluster
lensing and this effect is related to the 
{\it mass sheet degeneracy}].

\subsection{Galaxy Mass Distributions - parametric vs. non-parametric}

Parametric approaches have been favored from the beginning of lensing, as
it provides simple formulae and
gives analytical expressions of most or all the necessary lensing
quantities.
First, circular models were used like the point mass model; the singular 
isothermal sphere (SIS); the isothermal sphere with a core, with a truncation;
the NFW model (Navarro et al 1997); and recently more general cuspy models have been proposed (Holley-Bockelmann et al 2001).
The diversity of circular models has increased 
to allow more freedom in the radial profile of the mass distribution,
in particular following the recent developments of numerical simulations
of Dark Matter halos. By changing the slope of the mass profile
we can for example fix the image position but then allow a wider range
of acceptable time-delay or flux-ratio between multiple
 images as these quantities
will effectively depend on the radial profile.

Of course, circularity is not likely to be the property of galaxy 
mass distribution, hence the need of {\it elliptical mass distribution}.
However only the simplest mass profiles have an
analytical expression for the elliptical mass distribution. 
For the more complex ones
(such as the NFW model or the cuspy models), either {\it pseudo-elliptical}
models (Golse \& Kneib 2001)
or numerically integrated expressions have been proposed (Munoz et al 2001).

The interest of parametric expressions is their easy use, and their
predictive power. Furthermore, they are usually physically motivated 
and thus generally dynamically stable.

The non-parametric methods have been developed in the strong lensing regime
to allow more freedom in
the expression of the mass distribution. In general, the mass distribution
is represented as a pixelated array in the mass plane or alternatively
in the potential plane; the former being generally chosen as it
allows to use linear expressions (Saha \& Williams 1997).
Non-parametric 1D and 2D methods have been developed for cluster weak 
lensing, however only the 1D mass reconstruction from the weak shear profile
is of interest for galaxy lensing.
Although interesting, these approaches are poorly predictive and generally
proposed dynamically non-stable solutions.

Both methods should be explored as their results can be complementary,
however in the future we ought to develop multi-scale/multi-component
modeling which will take the best of the two current competing methods.

\section{Observations:\hfill -- Where Do We Fight --}

Better are the observations in terms of position, shape, time-delay,
better will be the constraints on the galaxy mass distribution that can
be derived. However, strong and weak lensing constraints usually do not 
overlap, in the sense that strong lensing focus on the 1 to 10 kpc
region, and the weak lensing on the 10 to 100 kpc region.

\subsection{Strong Lensing: Constraints and External Shear}
Because the number densities 
of background galaxies/quasars is small compared to the physical size
of a lensing galaxies, we currently know only a small number of
strong lensing systems. 

For a multiple quasar system, if $N$ is the number of multiply images 
observed, we thus have (as a maximum):
$2(N-1)$ constraints in position, $N-1$  constraints in amplification,
$N-1$ constraints in time-delay. For a double system, it means a maximum
of 4 constraints, but 12 for a quadruple system.
Of course, external constraints are usually used, like the observed
lensing galaxy center, its ellipticity and position angle (that is 
generally a maximum
of 4). This is to compare to the description of the mass distribution
of a galaxy which is represented at least by 6 or 7 parameters: 
$x, y, \varepsilon, \theta, \sigma_0, r_c, r_{cut}, \alpha$ ... 

One can see that for double quasar the number of constraints is of the order
of the number of free parameters, and that generally quadruple system are
over-constrained (assuming a one clump mass model).

Of course larger number of images can arise like the 6 image B1359+154
(Rusin et al 2001) or the 10 image radio-lens
B1933+503 (Cohn et al 2000), but generally they also need 
more complex mass distribution to
be properly understood.
An interesting avenue, is to detect the host galaxy of multiple quasars,
indeed, the larger the number of structures identified in the host galaxy the 
larger the number of constraints
on the mass distribution. If the host is sufficiently extended, then it will
form an Einstein ring as it has been recently observed and discussed by 
Kochanek et al (2001).  However, lensing constraints
are local and thus will only shade light on the mass distribution at the
location of the images. Thus ideally, we would like to observe multiple images
at different radius from the galaxy center, in order to probe accurately the 
mass {\it profile}.

The current situation has enable to show that rarely a 
perfect fit is obtained with
only one mass clump centered on the main lensing galaxy
({\it e.g.} Keeton, Kochanek, Seljak 1996). The simplest
way to improve the fit is to introduce what is called {\it external shear}:
a mathematical tweak of 2 parameters (the intensity $\gamma_E$ and its
orientation $\theta_E$). The only, but important drawback on 
the use of the {\it external shear} is that it is not physically motivated
has it has no mass. Understanding the origin of the external shear is 
currently an important GL question.

There is four possible origins to the external shear: 
1) the main galaxy itself (by allowing the radial mass profile to change 
for an elliptical mass distribution)
[as {\it e.g.} in HE2149-27, Burud et al 2001];
2) nearby galaxies [as in most systems?]; 
3) nearby group of galaxies [as {\it e.g.} in PG1115, in HST14176]; 
4) nearby cluster of galaxies [as {\it e.g.} in RXJ0911, Kneib et al 2001].

The external shear contribution increases with the size of the mass
perturbating the system, and decreases with its distance to the lensing system. 
Clearly, nearby
galaxies are the most likely origin, and clusters the less likely. However
because most of strong lensing galaxies are ellipticals the presence of
a nearby group or cluster is not surprising, as ellipticals are usually
found in dense environments.

To gain more information on the line of sight mass contribution, deep, 
wide multi-color images, followed by a redshift survey of the nearby structure,
or a deep X-ray observation, are needed to quantify precisely the 
different mass components and to explain the origin of external
shear.

\subsection{Galaxy-Galaxy Lensing: Scaling Laws and Recovery Methods}

As massive clusters in their outskirts, foreground
galaxies distort background galaxies 
following the weak lensing equation that reads 
(in the weak regime approximation):
$\vec\varepsilon_I = \vec\varepsilon_S + \vec{g}$ where 
$g={\gamma\over 1-\kappa}\sim \gamma$ is the reduced shear.
By averaging over the (a priori) random orientation of the sources the mean 
ellipticity of the images equals the reduced shear: 
$<\vec\varepsilon_I> = \vec{g}$, and the dispersion of the measurements
is $\sigma_g = {\sigma_{\varepsilon_S} \over \sqrt{N}}$. As 
$\sigma_{\varepsilon_S}\sim 0.25$, ideally, for a 4-sigma measure the 
number of galaxies needed scales as $N\sim g^{-2}$.
However due to the measurement errors (cirularisation and anisotropies of the
PSF --- see a number of contributions in this conference) the number of galaxies needed scales probably more as $N\sim 2-3 \times g^{-2}$.
As we are probing regions with $g\sim$ 0.01 to 0.0001 a very large amount
of quality data is required.

However this simple calculations over simplify the problem, indeed galaxies
have different sizes, luminosities, masses and are not at the same 
redshift. Using a simple approach, one can only constrain the
mass of the {\it average galaxy} which is not exactly what we aim to learn.
Therefore, it is important to use scaling laws and tune them
try to understand better the mass distribution of galaxies in their diversity.

The first critical scaling is the distance. Ideally, one wants to know the
redshifts of the lensing galaxies as they will allow to define their
angular diameter distances, and estimate their luminosities from their
broad band magnitudes. It is thus clear that any weak lensing survey with
multicolor and spectroscopic informations (for the brighter galaxies)
is of critical importance when we want to relate the galaxy-galaxy lensing
signal to galaxy mass distribution
({\it e.g.} the LCRS galaxy-galaxy lensing analysis: Smith et al 2001,
and the recent SDSS analysis: McKay et al 2001). 
Not having a spectroscopic redshift,
one can alternatively used photometry redshifts or any redshift informations
that can be derived from the broad-band photometry ({\it e.g.} the
early work of Brainerd et al 1996).

The second scaling is to assume that the mass distribution can be represented
by a {\it universal mass profile} which depends only on a very small number
of parameters. The general approach first proposed by Brainerd et al. (1996)
is to scale the velocity dispersion $\sigma$ and truncature radius $r_{cut}$
of an isothermal profile with the galaxy luminosity following some general
prescription like the Faber-Jackson law: $\sigma = \sigma_* (L/L_*)^{1/4}$,
the Kormendy relation: $r_{cut} = r_{{cut}^*} (L/L_*)^{0.8}$ [implying
$(M/L)\propto L^{0.3}$] or assuming $(M/L)=cste$ whatever the luminosity
[implying $r_{cut} = r_{{cut}^*} (L/L_*)^{0.5}$].
Of course, this is really the simplest one can assume, the mass distribution
may depend on the effective radius of the galaxy and/or its morphological
type, the exponent in the scaling relations may be different that the 
standard ones. Furthermore, the truncated isothermal
sphere may not represent the correct {\it universal mass profile}, 
hence other mass profile such as cuspy models should be investigated.

Finally, galaxy mass distribution is likely not circular. Thus, one
want to relate the ellipticity of the mass to the ellipticity of the light
(by reason of symmetry they should have the same orientation). Either
one can assume that mass and light have the same ellipticity or one can
try to understand what is the scaling law relating the 2 ellipticities.

The simplest recovering technique is what we can call the {\it direct averaging}
where we try to estimate the mass distribution directly from the 
PSF corrected measured ellipticities ({\it e.g.} Bridle et al 
in this conference proceeding). Basically this means we estimate the
absolute shear for a {\it galaxy pair} (the foreground and background galaxies
separated  by a distance $r$) at a scaled distance $r/r_s$ ($r_s$ is estimated
from the foreground galaxy properties, such has the half-light radius,
or a luminosity scaled radius) by  averaging 
background galaxy ellipticities:
$$ \bar\gamma(r/r_s) = < {D_{OS}\over D_{LS}} \epsilon_I(r)> $$
where the ratio of the angular distances corrects from the redshift
difference from one galaxy pair to another.

Although this {\it direct} technique is very simple and robust, it allows
only simple scaling for the mass, and the non-trivial contribution of galaxy 
clustering is directly
included in the results, giving more an estimated of the galaxy-mass
correlation function than the exact mass distribution of an average galaxy.
Furthermore, 1) the mass derived from the shear $\gamma$ suffers from the
so-called mass sheet degeneracy hence making difficult to derive any
absolute mass estimate; 
2) the direct average signal is washed out by any large scale mass distribution
and thus should not be applied directly in galaxy cluster fields
(Natarajan \& Kneib 1997).

The alternative to the {\it direct} approach, are {\it inverse} methods, 
such as the maximum likelihood
methods presented in Schneider \& Rix (1996) and Natarajan \& Kneib (1997).
In these methods, we consider each (background) galaxy $i$ lensed by the nearby
(foreground) galaxies. Assuming some scaling laws (see above) one can
predicts the expected induced distortion on each (background) galaxy
and thus compute its intrinsic ellipticity $\vec\epsilon_{Si}$. 
Then by maximizing the 
likelihood $L = \Pi p_S(\vec\epsilon_{Si})$, where $p_S$ is the unlensed
galaxy ellipticity distribution, one will be able to derive the best
model that fits the observed data.

This strategy, although complex, allows to 
1) probe various scaling laws, 
2) test different form for the mass distribution profile,
3) use elliptical mass distribution,
4) model higher density environments like groups or clusters;
thus this is the one to select specially whit good quality data 
which is likely to be the case for current and future surveys.

\subsection{Strong and Weak Lensing: Galaxy in Clusters and Mass Evolution}

It has been realized ({\it e.g.} Kneib et al 1996)
that is is compulsory to take into account the mass distribution of galaxies
in clusters to accurately model the lensing distortion. In fact as shown
by Natarajan \& Kneib (1997) the presence of a large scale mass distribution
boost the galaxy-galaxy signal making it easier to detect if one used
an adequate method. This is opening prospects to try to understand
how the mass distribution is (re)distributed from small scale to
large scale as a function of time and local density. Such results will
be of great interest and will be important to compare to numerical
simulations. Such results are just coming along (Natarajan et al 2000)
and are likely to be of great interest with the development of
weak lensing surveys.

\section{The Future: \hfill -- How Will We Do --}

To better understand the higher mass densities of galaxies, we will need
to enlarge the number multiple image systems. This will come either
by current facilities - for example searching for small separation multiple
quasars in the new quasar surveys (SDSS, 2dF) or by future surveys
(ACS/SNAP/NGST/radio). When we have increased the number of systems
from the current $\sim 30$ to more than one thousand, we should be able to
probe accurately the galaxy mass distribution 
{\it vs.} galaxy type, environment and redshift.
We also need to better constrain the current multiple image systems, this
is possible by probing more accurately the line-of-sight mass distribution
(origin of the external shear, measure of the time-delay, accurate redshifts)
and by applying strong+weak lensing techniques.

Galaxy-galaxy lensing is likely to become sort of an {\it industry} with
the developments of high quality imaging and spectroscopic surveys and
will allow to test the various scaling laws and possible {\it universal}
mass distribution.
When applied to cluster survey, galaxy-galaxy lensing will allow to test
the stripping efficiency on galaxy scale in higher densities environment.
A possible interesting avenue, will also to conduct quasar-galaxy 
lensing survey to probe the weight of QSOs and their hosts and compare
this results to {\it normal} galaxy mass distribution.

In short, there are good prospects to learn more on galaxy mass distribution 
(baryonic and dark matter) in the near future!

\section*{Acknowledgments}
I thank Priya and the LOC for organizing this very useful conference.
I acknowledge support from INSU-CNRS, and from EU through the research
network {\it LENSNET} [http://webast.ast.obs-mip.fr/lensnet].

\end{document}